\documentclass[pra,aps,twocolumn,superscriptaddress,showpacs]{revtex4}

\usepackage{bm}

\newcommand{\beq}{\begin{equation}}
\newcommand{\eeq}{\end{equation}}
\newcommand{\beqa}{\begin{eqnarray}}
\newcommand{\eeqa}{\end{eqnarray}}
\newcommand{\ket} [1] {\vert #1 \rangle}
\newcommand{\bra} [1] {\langle #1 \vert}

\begin{document}
\title{Economical quantum cloning in any dimension}

\author{Thomas Durt}
\affiliation{TONA-TENA Free University of
Brussels, Pleinlaan 2, B-1050 Brussels, Belgium}

\author{Jarom\'{\i}r Fiur{\'a}{\v s}ek}
\affiliation{QUIC, Ecole Polytechnique, CP 165,
Universit\'{e} Libre de Bruxelles, 1050 Brussels, Belgium}
\affiliation{Department of Optics, Palack\'{y} University,
17. listopadu 50, 77200 Olomouc, Czech Republic}

\author{Nicolas J. Cerf}
\affiliation{QUIC, Ecole Polytechnique, CP 165,
Universit\'{e} Libre de Bruxelles, 1050 Brussels, Belgium}

\begin{abstract}
The possibility of cloning a $d$-dimensional quantum system
without an ancilla is explored, extending on 
the economical phase-covariant cloning machine found
in [Phys. Rev. A {\bf 60}, 2764 (1999)] for qubits.
We prove the impossibility of constructing an economical version of
the optimal universal cloning machine in any dimension. We also show,
using an ansatz on the generic form of cloning machines,
that the $d$-dimensional phase-covariant cloner, which optimally
clones all uniform superpositions, can be realized economically 
only in dimension $d=2$. The used ansatz is supported by numerical 
evidence up to $d=7$. An economical phase-covariant cloner 
can nevertheless be constructed for $d>2$, albeit with a lower fidelity than that of the optimal cloner requiring an ancilla. 
Finally, using again an ansatz on cloning machines, we show
that an economical version of the Fourier-covariant cloner,
which optimally clones the computational basis and its Fourier transform,
is also possible only in dimension $d=2$.
\end{abstract}


\maketitle

\section{Introduction}

During the last decade, many promising applications of ideas developed
within the framework of quantum information theory, such as
quantum cryptography, quantum computing, quantum cloning, and
quantum teleportation were implemented experimentally
\cite{RMP,Cummins,tele,du,Gulde}. Although it is not certain whether
these spectacular progresses will lead to a practical
quantum computer \cite{Jones} because of the difficulties related
to decoherence, quantum cryptography is already a well established
and mature technology \cite{RMP,naturecerf}. Traditionally, quantum key
distribution is implemented with two-level quantum systems, usually referred
to as qubits. The security of the quantum key distribution (QKD) protocols
such as the BB84 protocol \cite{BB84} is guaranteed by the {\it no-cloning}
theorem \cite{Dieks,Wootters}, which states that the perfect copying
(or cloning) of a set of states that contains at least two non-orthogonal
states is impossible. It is, however, possible to
realize an approximate quantum cloning, a concept introduced in a
seminal paper by Bu\v{z}ek and Hillery \cite{buzek} where a universal
(or state-independent) and symmetric one-to-two cloning
transformation was introduced for qubits.

The cloning machines can be used as very efficient eavesdropping attacks
on the QKD protocols. In this context, it is important to study
machines which optimally clone a particular subset of states of the Hilbert
space, for example the Fourier-covariant cloning machine, which optimally
copies two mutually unbiased bases under a Fourier transform \cite{bourennane}, or the phase-covariant cloning machine, 
which optimally clones all balanced superpositions
of the computational basis states \cite{NG,bruss,cerfdurtgisin,Fan02,fan,Rezakhani03,kwek,Lamoureux04}.
In particular, the optimal Fourier-covariant cloner in two dimensions, 
is known to provide the most dangerous eavesdropping strategy
for the BB84 quantum cryptographic protocol \cite{BB84}, 
while the phase-covariant and universal cloners respectively 
play the same role relatively to the Ekert \cite{Ekert} and 6-states \cite{bruss6states,bechmann-gisin} protocols.

In the present paper, we shall concentrate on the
one-to-two cloning machines, which produces two copies. 
In an eavesdropping scenario, one copy is sent to the legitimate receiver 
while the other one is kept by the eavesdropper.
The $1\rightarrow 2$ cloning transformation for qudits can typically be
expressed as a unitary operation on the Hilbert space of three qudits
--- the input, a blank copy, and an ancilla. The presence of ancilla
significantly affects the experimental implementation
of the cloning operation, which becomes more complicated and sensitive to
decoherence as it has been shown in a recent NMR realization of optimal
universal qubit cloner \cite{Cummins}. These negative effects,
which may drastically reduce the achieved cloning fidelity, 
may significantly be suppressed if an ``economical''approach is followed,
which avoids the ancilla. The cloning is then realized 
as a unitary operation on two qudits only: the
input and the blank copy. This is obviously
much simpler to implement because it requires less qudits and two-qudit gates, and it requires to control the entanglement
of a pair of qudits only. It is thus likely to be much less sensitive to
noise and decoherence than its three qudit counterpart, a fact that
was recently confirmed experimentally \cite{expe}. To date, the only
$1 \rightarrow 2$ cloning machine for which an economical realization 
is known is the phase-covariant qubit cloner due to
Niu and Griffiths \cite{FGGNP,NG,lowcostcloning}.

The (asymmetric) phase-covariant qubit cloning machine \cite{NG} 
works as follows. During the process,
the qubit to be cloned, initially in state $\ket{\psi}_{B}$,
is coupled to another qubit which become the second copy
and is initially prepared in state $\ket{0}_{E}$ 
(the labels $B$ and $E$ refer to the tradition in quantum cryptography
according to which the receiver of the key
is called Bob and the eavesdropper Eve). Then, the state
$\ket{\psi}_{B}\ket{0}_{E}$ undergoes a unitary transformation $
U_{BE}$ such that
\beqa 
U_{BE}\ket{0}_{B}\ket{0}_{E}&=&\ket{0}_{B}\ket{0}_{E}
\nonumber \\
U_{BE}\ket{1}_{B}\ket{0}_{E}&=&\cos\alpha \,\ket{1}_{B}\ket{0}_{E} +
\sin\alpha \,\ket{0}_{B}\ket{1}_{E}
\label{NGtransf}
\eeqa 
It can be shown that when the input qubit is
in an equatorial state,
\beq
\ket{\psi}_{B}={1\over \sqrt 2}
\left(\ket{0}_{B}+e^{i\phi}\ket{1}_{B}\right)
\eeq
the fidelities of Bob's and Eve's clones give
\beqa
F_B&=&\bra{\psi}_{B}\text{Tr}_{E}(\rho)\ket{\psi}_{B}
={1+\cos\alpha\over 2}  \nonumber \\
F_E&=&\bra{\psi}_{E}\text{Tr}_{B}(\rho)\ket{\psi}_{E}
={1+\sin\alpha\over 2}
\eeqa
where $\rho=|\Phi_{BE}\rangle\langle\Phi_{BE}|$
and $|\Phi_{BE}=U_{BE}\ket{\psi}_{B}\ket{0}_{E}$
These fidelities do not depend on the azimuthal angle $\phi$,
so that these cloners are called phase-covariant.
The special case $\alpha=\pi/4$ corresponds to the symmetric 
phase-covariant cloner, which provides two clones 
of equal fidelity $F_B=F_E=(2+\sqrt 2)/4\approx 0.85$.


It is worth emphasizing that, excepted for the two qubits which are used
to carry the two copies, this transformation does not require any extra qubit (ancilla), and is thus an economical cloning process. In a recent paper,
a general, necessary and sufficient, criterion was derived in order to
characterize the reducibility of 3-qubit cloners to 2-qubit cloners, and
it was concluded that the phase-covariant cloner is the only cloner
in dimension $d=2$ that admits an economical realization \cite{lowcostcloning}.
The goal of the present paper is to further extend this study,
and to investigate whether a two-qudit
realization exists also for $d$-dimensional cloning machines.
More generally, we aim at elucidating 
the connections that exist between the cloners with or without ancillas.
We prove a series of no-go theorems for economical one-to-two cloning. In
particular, we show that, without an ancilla,
it is impossible to realize the (deterministic) optimal universal cloning machine in any dimension $d$ (Section II),
and that an economical implementation of optimal phase-covariant cloners
does not exist for dimensions $d>2$ (Section III). This latter result 
relies on some ansatz on the cloning transformation, which is made
very plausible by a numerical check up to $d=7$. As a side-result,
we also consider the best economical phase-covariant cloner in $d$ dimensions,
which achieves a high fidelity although it does not perform as well
as the optimal phase-covariant cloner with an ancilla (Section IV).
Moreover, we provide a strong evidence that the
optimal cloning of a pair of mutually unbiased bases, or Fourier-covariant
cloning, requires an ancilla if $d>2$ (Section V). All these results
strongly suggest that the Niu-Griffiths phase-covariant qubit cloner
\cite{NG}, which does not require an ancilla, is quite unique among the
$1 \rightarrow 2$ cloning machines.

\section{Universal cloning machines}

Let us begin by introducing an isomorphism between completely positive maps
$\mathcal{S}$ and positive semidefinite operators $S \geq 0$ on the tensor
product of input and output Hilbert spaces, $\mathcal{H}_{\mathrm{in}}\otimes
\mathcal{H}_{\mathrm{out}}$ \cite{Jamiolkowski72,Choi75}. Consider a maximally entangled
state on $\mathcal{H}_{\mathrm{in}}^{\otimes 2}$,
\begin{equation}
|\Phi^+\rangle=\frac{1}{\sqrt{d}}\sum_{j=1}^d |j\rangle_{1}|j\rangle_{2},
\end{equation}
where $d=\mathrm{dim}(\mathcal{H}_{\mathrm{in}})$
The map $\mathcal{S}$ is applied to the subsystem $2$, while nothing happens
with subsystem 1. The resulting (generally mixed) quantum state is isomorphic
to $\mathcal{S}$ and reads
\begin{equation}\label{S3}
S= \mathcal{I}_{1}\otimes \mathcal{S}_2 (d |\Phi^+\rangle\langle \Phi^+|).
\end{equation}
The prefactor $d$ is introduced for normalization purposes.
A trace preserving map satisfies the condition
\begin{equation}
\mathrm{Tr}_{\mathrm{out}} [S]=\openone_{\mathrm{in}}.
\label{tracepreservation}
\end{equation}
The CP map $\rho_{\mathrm{out}}=\mathcal{S}(\rho_{\mathrm{in}})$ can be expressed in terms of $S$
as follows \cite{Fiurasek01},
\begin{equation}
\rho_{\mathrm{out}}= \mathrm{Tr}_{\mathrm{in}}[\rho_{\mathrm{in}}^T \otimes \openone_{\mathrm{out}} S],
\end{equation}
where $T$ denotes the transposition in the computational basis.

Let us now consider that $S$ describes the $1\rightarrow 2$ cloning
transformation of qudits. The output Hilbert space is endowed with tensor
product structure, $\mathcal{H}_{\mathrm{out}}=\mathcal{H}_{B}\otimes\mathcal{H}_E$,
where the subscripts $B$ and $E$ label the two clones (in the framework of quantum cryptography, they label the authorized user's (Bob's) copy and the spy's (Eve's) copy). For each particular
input state $|\psi\rangle$, we can calculate the fidelity of each clone as
follows,
\begin{eqnarray}
F_{B}(\psi)&=&\mathrm{Tr}(\psi_{\mathrm{in}}^T\otimes \psi_{B}\otimes \openone_E S),
\nonumber \\
F_{E}(\psi)&=&\mathrm{Tr}(\psi_{\mathrm{in}}^T\otimes \openone_{B}\otimes \psi_E S),
\label{FBEpsi}
\end{eqnarray}
where $\mathrm{in}$ labels the input and $\psi\equiv |\psi\rangle\langle \psi|$ is a short
hand notation for a density matrix of a pure state. We are usually interested
in the average performance of the cloning machine, which can be quantified
by the
mean fidelities,
\begin{equation}
F_B =\int_{\psi} F_B(\psi) d\psi, \qquad
F_E =\int_{\psi} F_E(\psi) d\psi,
\label{FAB}
\end{equation}
where the measure $d\psi$ determines the kind of the cloning machines we are
dealing with. Universal cloning machines correspond to choosing $d\psi$ to be
the invariant measure on the factor space $SU(d)/SU(d-1)$ induced by the Haar
measure on the group $SU(d)$. The fidelities (\ref{FAB}) are linear
functions of the operator $S$,
\begin{equation}
F_{B}=\mathrm{Tr}[SR_{B}], \qquad F_{E}=\mathrm{Tr}[S R_E],
\end{equation}
where the positive semidefinite operators $R_j$ are given by
\begin{equation}
R_{B}=\int_{\psi} \psi_{\mathrm{in}}^T \otimes \psi_B \otimes \openone_E d \psi,
\qquad
R_{E}=\int_{\psi} \psi_{\mathrm{in}}^T \otimes \openone_B \otimes \psi_E d \psi.
\label{RBEdefinition}
\end{equation}
In case of universal cloning, the integral over $d\psi$ can be easily calculated
with the help of Schur's lemma, and we get, for instance,
\begin{eqnarray*}
\int_{\psi} \psi_{\mathrm{in}}^T \otimes \psi_{B} d \psi&=&
\frac{2}{d(d+1)}(\Pi_{\mathrm{in},B}^{+})^{T_{\mathrm{in}}} \\
&=&\frac{1}{d(d+1)}[\openone_{\mathrm{in}}\otimes\openone_{B}+d\, \Phi_{\mathrm{\mathrm{in}},B}^{+}].
\end{eqnarray*}
Here, $\Pi^{+}$ denotes a projector onto symmetric subspace of two qudits,
$d(d+1)/2$ is the dimension of this subspace, and $T_{\mathrm{in}}$ stands for transposition
with respect to the subsystem $\mathrm{in}$.

The optimal symmetric cloning machine $S$ should maximize the
average of mean fidelities $F_{B}$ and $F_{E}$ \cite{Fiurasek03},
\begin{equation}
F=\frac{1}{2}(F_{B}+F_{E})=\mathrm{Tr}[SR],
\end{equation}
where $R=(R_B+R_{E})/2$.
The maximum achievable $F$ is upper bounded by the maximum eigenvalue
$r_{\mathrm{\mathrm{max}}}$ of the operator $R$. Taking into account the
trace-preservation condition (\ref{tracepreservation}),
we have \cite{Fiurasek01}
\begin{equation}\label{www}
F \leq d r_{\mathrm{max}}.
\end{equation}
In the case of the universal and phase-covariant $1\rightarrow 2$ cloning
machines considered in the present paper this bound is saturated if we use an
ancilla as we shall see below.

We have to calculate the eigenvalues of an operator
\begin{equation}
R=\frac{1}{2d(d+1)}(2\openone_{\mathrm{in}BE}+d \Phi_{\mathrm{in},B}^{+}\otimes \openone_{E}+
d\Phi_{\mathrm{in},E}^{+} \otimes \openone_B),
\end{equation}
Due to the high symmetry, the operator $R$ has only
there different eigenvalues. One eigenvalue reads $1/(d(d+1))$ and
is $d^3-2d$-fold degenerate. The other two eigenvalues are each $d$-fold
degenerate and the corresponding eigenstates lie in the $2d$-dimensional
subspace spanned by $|\Phi^+\rangle_{\mathrm{\mathrm{in}},B}|k\rangle_E$ and
$|\Phi^+\rangle_{\mathrm{\mathrm{in}},E}|k\rangle_B$, with $k=1,\ldots,d$.
The $d$ eigenstates corresponding to the maximum eigenvalue read,
\begin{equation}\label{yyy}
|r_{\mathrm{max}};k\rangle=\sqrt{\frac{d}{2(d+1)}}
(|k\rangle_{B}|\Phi^+\rangle_{\mathrm{\mathrm{in}},E}
+|k\rangle_{E}|\Phi^+\rangle_{\mathrm{\mathrm{in}},B}),
\label{rtilde}
\end{equation}
where $k=1,\ldots,d$.
It is clear that the support of any admissible optimal cloning CP
map $S$ must be the $d$-dimensional space spanned by the eigenstates
$|r_{\mathrm{max}};k\rangle$. This will be exploited in what follows to prove that
it is not possible to implement the cloning transformation in an economic way,
i.e. without an ancilla, just by applying (randomly, with probability $p_l$)
a two-qudit unitary transformation $U_l$ to the original state and a blank copy.

If this convex mixture of the unitaries implements optimal
cloning transformation which maximizes the fidelity $F$, then, by convexity,
each unitary $U_l$ is optimal in a sense that it yields the maximal mean
fidelity. Consider one such unitary $U$. The corresponding operator $S_U$ represents
a pure state, since $S_U$ is obtained by applying $U$ to a pure state
$|\Phi^{+}\rangle$. The question is thus whether there exists a state
\begin{equation}
|S_U\rangle=\sum_{k=1}^dc_{k}|r_{\mathrm{max}};k\rangle
\end{equation}
such that $S_U=|S_U\rangle\langle S_U|$ satisfies the trace-preservation
condition (\ref{tracepreservation}). After a simple algebra, the condition
$\mathrm{Tr}_{BE}[S_U]=\openone_{\mathrm{in}}$ turns out to be equivalent to
\begin{equation}
\frac{1}{d+1} \sum_{k=1}^d|c_{k}|^2 \openone
+\frac{1}{d+1}\sum_{k,l} c_k c_l^\ast
|l\rangle\langle k| =\openone.
\end{equation}
This condition is equivalent to the requirement that the rank-one projector
$|c^{\ast}\rangle\langle c^{\ast}|$ is proportional to the identity operator,
which is clearly impossible for any dimension $d \geq 2$. This concludes our
proof that the universal $1\rightarrow 2$ economical cloning is impossible.

\section{Phase-covariant cloning machines}

Let us now investigate the possibility of the economical implementation of
phase-covariant cloning machines which clone equally well all
balanced superpositions of computational basis states,
\[
|\psi\rangle= \frac{1}{\sqrt{d}} \sum_{j=1}^d e^{i\phi_j} |j\rangle.
\]
We will proceed similarly as before and first determine the operators
$R_B^{pc}$ and $R_E^{pc}$, where the superscript pc indicates states and
operators related to phase-covariant cloning. The integration in
Eq. (\ref{RBEdefinition}) is over the $d$ phases $\phi_j$, and we have
to evaluate the integral
\begin{eqnarray*}
\prod_{j=1}^d \int_{0}^{2\pi} \frac{d\phi_j}{2\pi} \psi_{\mathrm{in}}^T\otimes
\psi_B
&=&\frac{1}{d}
\Phi_{\mathrm{\mathrm{in}},B}^{+} + \frac{1}{d^2}\openone_{\mathrm{in}} \otimes\openone_B
\nonumber \\
&&-\frac{1}{d^2}\sum_{j=1}^d (|jj\rangle\langle jj|)_{\mathrm{\mathrm{in}},B} .
\end{eqnarray*}
In order to determine the subspace that is the support of all possible optimal
cloning transformations $S^{pc}$, we have to determine the maximum eigenvalue of the
operator $R^{pc}=(R_B^{pc}+R_E^{pc})/2$ and the corresponding eigenstates. We have
\begin{eqnarray}
R^{pc}&=&\frac{1}{d^2}
\openone_{\mathrm{in}}\otimes\openone_B\otimes\openone_E
\nonumber \\
& &+\frac{1}{2d}\left(\Phi_{\mathrm{\mathrm{in}},B}^{+}\otimes \openone_E
+ \Phi_{\mathrm{in,E}}^{+}\otimes\openone_{B}\right)
\nonumber \\
& & -\frac{1}{2d^2}\sum_{j=1}^{d}\left[(|jj\rangle\langle jj|)_{\mathrm{\mathrm{in}},B}\otimes\openone_E
+(|jj\rangle\langle jj|)_{\mathrm{\mathrm{in}},E}\otimes\openone_B \right].
\nonumber \\
\end{eqnarray}
Taking into account the symmetry properties of the operator $R^{pc}$, we can make an ansatz
for the eigenstates of $R^{pc}$ which correspond to the maximum eigenvalue,
\begin{equation}\label{zzz}
|r_{\mathrm{max}}^{pc};k\rangle=\alpha(|\Phi^+\rangle_{\mathrm{\mathrm{in}},B}|k\rangle_B+|\Phi^+\rangle_{\mathrm{\mathrm{in}},E}|k\rangle_B)
+\beta|kkk\rangle_{\mathrm{in},BE},
\label{rmaxpc}
\end{equation}
where $k=1,\ldots,d$, and
\begin{equation}
\frac{\alpha}{\beta}=-\frac{\sqrt{d}}{4}(d+2+\sqrt{d^2+4d-4}).
\label{alphabetapc}
\end{equation}
One can easily verify that $|r_{\mathrm{max}}^{pc};k\rangle$ is indeed
an eigenstate of $R^{pc}$ if the condition (\ref{alphabetapc}) is satisfied.
However, it is much more difficult to prove that it is the eigenstate with
highest eigenvalue and that the $d$ states (\ref{rmaxpc}) are the \emph{only}
eigenstates with this maximum eigenvalue. While we have not been able to prove this
analytically for arbitrary $d$, we have checked numerically that this is indeed
the case for $d=2,3,\ldots,7$ and we conjecture that this holds for any
$d$.

We can now prove that for $d>2$ it is not possible to design an economical
phase-covariant cloning machine which does not require an ancilla.
If such a machine would exist, then there would be a state
\begin{equation}
|S^{pc}\rangle=\sum_{k=1}^d c_{k} |r_{\mathrm{max}}^{pc};k\rangle,
\label{Spctracepreservation}
\end{equation}
which would satisfy the trace-preservation condition (\ref{tracepreservation}).
On inserting (\ref{rmaxpc}) into (\ref{Spctracepreservation}) we obtain
\begin{eqnarray}
\mathrm{Tr}_{BE}(|S^{pc}\rangle\langle S^{pc}|)&=&2d^{-1}\sum_k |c_k|^2 \openone
+\gamma \sum_{k} |c_k|^2 |k\rangle\langle k|
\nonumber \\
&&+2\frac{\alpha^2}{d} \sum_{j \neq k} c_k c_j^\ast |j\rangle \langle k|,
\nonumber \\
\end{eqnarray}
where
\[
\gamma=\beta^2+\frac{4\alpha\beta}{\sqrt{d}}+\frac{2\alpha^2}{d}.
\]
We have to distinguish two cases. If $\gamma=0$ then the trace-preservation
condition (\ref{tracepreservation}) can be satisfied by setting
$c_k=0$ if $k \neq l$ and $c_l=\sqrt{d/2}$ for some $l \in \{1,\ldots ,d\}$.
From $\gamma=0$ we obtain $\alpha/\beta=-\sqrt{d}(4\pm 2\sqrt{2})/4.$
By comparing this expression with Eq. (\ref{alphabetapc}) we obtain an
equation for $d$ which has only one positive integer solution $d=2$.
In this particular case, the pure state $|r_{\mathrm{max}}^{pc};k\rangle$
describes the symmetric Niu-Griffiths phase-covariant
cloning machine for qubits \cite{NG} and we have, in accordance with Eqs..(\ref{NGtransf},\ref{S3}),
\[
|S^{pc}\rangle=|0\rangle_{\mathrm{in}}|00\rangle_{BE}
+\frac{1}{\sqrt{2}}|1\rangle_{\mathrm{in}}(|01\rangle+|10\rangle)_{BE}.
\]
For $d>2$ it holds that $\gamma \neq 0$ and
the trace-preservation condition thus implies $c_k c_j^\ast = C \delta_{jk}$,
where $C>0$ is some constant. It is clear that this latter constraint
does not admit any solution, hence we conclude that for $d>2$ the economical
phase-covariant cloning machine does not exist. Strictly speaking, our proof
holds only for $d=3,\ldots, 7$ where we numerically verified that the
eigenstates (\ref{rmaxpc}) are the only ones corresponding to the maximal
eigenvalue of $R^{pc}$, however, we expect that it holds for any $d>2$.

\section{Suboptimal economical phase-covariant cloning machines}

Since the optimal phase-covariant cloning cannot be realized without an
ancilla, we can ask what is the best economical approximation to the optimal
cloner, i.e., which unitary operation on the Hilbert space of two
qudits, an input and a blank copy, achieves the maximum cloning fidelity.
In our formalism, the unitary operation is represented by a rank one operator
$S_U=|S_U\rangle \langle S_U|$ which satisfies
$\mathrm{Tr}_{AB}[|S_U\rangle\langle S_U|]=\openone_{\mathrm{in}}$.
The optimal $U$ can be easily determined if we impose some natural constraints
on the cloning transformation. First of all, we require that it should be
invariant with respect to swapping the two clones $A$ and $B$, which implies
that the output Hilbert space of $S_U$ should be the symmetric subspace of the two
qudits, spanned by the states $|kl^{+}\rangle$ defined as
$|kl^{+}\rangle=(|kl\rangle+|lk\rangle)/\sqrt{2}$, $k \neq l$, and
$|kk^{+}\rangle=|kk\rangle$. The second condition is that the cloning should be
phase covariant, i.e. the map $S_U$ should be invariant with respect to an
arbitrary phase shift applied to the input qubit, followed by the inverse phase
shifts on the two clones. Mathematically, this condition can be expressed as
\begin{equation}
[V_{\mathrm{in}}(\bm{\phi}) \otimes V_{B}^{\dagger}(\bm{\phi})\otimes
V_E^{\dagger}(\bm{\phi})]|S_U\rangle= e^{i\phi} |S_U\rangle,
\label{phasecovariance}
\end{equation}
where $\phi$ is some overall phase factor,
\[
V(\bm{\phi})=\sum_{k=1}^d e^{i \phi_{k}}|k\rangle\langle k|,
\]
and the phases $\phi_k$ can be arbitrary. In order to satisfy the condition
(\ref{phasecovariance}), the state $|S_{U}\rangle$ must have one of the following forms
\begin{eqnarray}
|S_U\rangle&=&|k\rangle_{\mathrm{in}}|lm^{+}\rangle_{BE}, \qquad k \neq l \neq m,
\nonumber \\[2mm]
|S_U\rangle&=&|k\rangle_{\mathrm{in}}|ll^{+}\rangle_{BE}, \qquad k \neq l ,
\nonumber \\
|S_U\rangle&=&\sum_{k=1}^d s_k |k\rangle_{\mathrm{in}}|kl^{+}\rangle_{BE}.
\nonumber
\end{eqnarray}
It is clear that the trace preservation condition can be satisfied only by the
third option, provided that $s_k= e^{i \theta_k}$. The fidelity of the clones
produced by this map is given by
\[
F=\frac{1}{2d^2}(d-1+|\sum_{k \neq l} e^{i\theta_k}+\sqrt{2}e^{i\theta_l}|^2)
\]
and is maximized when $\theta_k=0$, $k=1,\ldots,d$.The optimal economical
phase-covariant cloning transformation which is invariant with respect to the
swapping of the two clones and is also phase covariant can be thus expressed as
\[
|k\rangle \rightarrow |kl^{+}\rangle,
\]
where $l \in \{1,\ldots,d,\}$ is arbitrary, and the corresponding
fidelity reads
\[
F_U=\frac{1}{2 d^2}[d-1+(d-1+\sqrt{2})^2].
\]

\section{Fourier-covariant cloning machines}

Although it is not always easy to prove analytically or numerically that
certain cloning machines optimize given quantities (like Bob and Eve's
fidelities), an educated guess is often possible. For instance, one can show
that the overwhelming majority of optimal 1 to 2 cloning machines that can
be found in the literature obeys \cite{inprep} the ansatz given in Refs. \cite{CERFPRL,CERF}.
According to this ansatz, the cloning transformation is represented by
a pure state in a $d^4$ dimensional space spanned by the qudits conventionally
labeled by $A$, $B$, $E$ and $M$ where $A$ represent Alice's qudit and
is formally equivalent to the label $in$ introduced in the previous section,
$B$ and $E$ represent Bob's and Eve's qudits as before, while $M$ represents
an external ancilla. Moreover, the cloning state is assumed to be biorthogonal
in the Bell bases, where the $d^2$ qudit Bell states are defined as follows:
\beq
\label{Bell}
\ket{B_{m,n}}_{1,2}={1\over\sqrt{d}}\sum_{k=0}^{d-1}
\gamma^{kn}\ket{k}_{1}\ket{k+m}_{2}
\eeq
where $m,n \in \{0,1,...,d-1\}$, $\gamma$ is the $d$-th root of unity,
and $\ket{j}_{1(2)}$ represents a state of the qudit
system $1$ ($2$) chosen in the computational basis.
They are maximally entangled states and form an orthonormal basis of
the $d^2$-dimensional Hilbert space of qudits $1$ and $2$.
Because the cloning state is biorthogonal in the Bell bases, it can
be expressed as follows:
\begin{equation}
\label{ansatz}
\ket{\Psi}_{A,B,E,M}=\sum_{m,n=0}^{d-1}
a_{m,n}\ket{B_{m,n}}_{A,B}\ket{B_{m,-n}}_{E,M}
\end{equation}
Here $a_{m,n}$ is a (normalized) $d \times d$ matrix. The specification of
the $d^2$ amplitudes $a_{m,n}$ defines the cloning transformation.
We now give several examples.

The optimal universal (generally asymmetric) cloning machine is
defined by the following amplitude matrix,
\beq
\label{univ}
a^U_{m,n}=x_1\delta_{m,0}\delta_{n,0}+x_3
\eeq
The optimal {\em symmetric} universal $d$-dimensional cloner
(the one for which Eve's fidelity is maximal, under the constraint that
Bob's fidelity is equal to Eve's fidelity) is obtained by choosing
$x_1^2=x_3^2=d/[2(d+1)]$. It copies all states with the same fidelity,
and we recover the standard formula for the fidelity of universal cloners
\cite{buzek98,werner,CERFPRL,CERF} $F=(3+d)/[2(1+d)]$.

The qubit phase covariant cloner copies equally well two mutually unbiased
qubit bases (maximally-conjugate or mutually unbiased bases are such that
any basis state in one basis has equal squared amplitudes when expressed
in any other basis). As far as we presently know, the most dangerous
attack on the BB84 \cite{BB84} and Ekert's \cite{Ekert}
protocols requires Eve to make use of such a cloner.
It possesses two interesting generalizations in higher dimension:
(a) the phase-covariant cloner and (b) the Fourier-covariant cloner.

(a) The phase-covariant cloner has already been defined in the previous
section; it clones equally well all balanced superpositions of computational
basis states, $|\psi\rangle= \frac{1}{\sqrt{d}} \sum_{j=1}^d e^{i\phi_j} |j\rangle$. The asymmetric phase-covariant cloning machine is described (for arbitrary dimension)
in Ref.~\cite{kwek} (and the symmetric one in \cite{fan}).
It is defined by the following amplitude matrix:
\begin{equation}
\label{phasecov}
a^{PC}_{m,n}=x_{1}\delta_{m,0}\delta_{n,0}+x_{2}\delta_{m,0}+x_{3}
\end{equation}
where $x_{1}$, $x_{2}$ and $x_{3}$ are real positive parameters.
It constitutes the most dangerous currently known attack on $d$-dimensional
generalizations of Ekert's protocol.

(b) The Fourier-covariant cloner clones equally well two mutually unbiased
bases that are discrete Fourier transforms of each other \cite{bourennane}.
It constitutes the most dangerous attack
on d-dimensional generalizations of the BB84 protocol.
The Fourier cloner is characterized
by the following amplitude matrix \cite{nagler},
\begin{equation}
\label{Fouriercov'}
a^F_{m,n}=x_{1}\delta_{m,0}\delta_{n,0}+x_{2}
(\delta_{m,0}+\delta_{n,0})+x_{3},
\end{equation}
where $x_{1}$, $x_{2}$ and $x_{3}$ are real positive parameters.


It is legitimate
to ask whether or not an economic realization of such an optimal cloning machine
is possible, so to say whether it is possible to reach the same fidelity
without making use of the ancilla. Concretely, this means that it is possible
to find $l_{\mathrm{max}}$ probabilities $p_{l}$ and $l_{\mathrm{max}}$ unitary
transformations $U^l_{BE}$ that act on the qudits $B$ and $E$ only such that:
\begin{equation}
\label{lowcostoptimal}
S_{ABE}=\mathrm{Tr}_{M}\Psi^{opt}_{A,B,E,M} \nonumber \\
=\sum_{l=1...l_{\mathrm{max}}} p_{l} \Phi^{l}_{A,B,E} ,
\end{equation}
where $\Psi=|\Psi\rangle\langle\Psi|$ and $\Phi=|\Phi\rangle\langle\Phi|$
are short-hand notations for density matrices of pure states, and
\begin{equation}
\label{lowcostoptimali}
\ket{\Phi}^{l}_{A,B,E}={1 \over \sqrt d }\sum_{k=0}^{d-1}
\ket{k}_{A}U^l_{BE}\ket{k}_{B}\ket{\psi_{0}}_{E}.
\end{equation}
As a consequence of the convexity of the average fidelity of cloning,
if the CP map $S_{ABE}$ represent an optimal cloning transformation
then each unitary transformation $\Phi_{A,B,E}^l$ is also optimal in a sense
that it maximizes the average cloning fidelity.
The support of the CP map $S$ associated with
the cloning machines that fulfill the ansatz (\ref{ansatz}) is spanned
by the $d$ states
\begin{equation}
|r_p\rangle= _M\bra{p} \sum_{m,n=0}^{d-1}a_{m,n}
\ket{B_{m,n}}_{A,B}\ket{B_{m,-n}}_{E,M},
\end{equation}
where $p \in \{0,1,...,d-1\}$. In what follows we assume that the states
$|\xi_p\rangle$ are eigenstates with maximum eigenvalue $r_{\mathrm{max}}$
of an operator $R$ which appears in the formula for the cloning fidelity,
$F=\mathrm{Tr}[R S_{ABE}]$.
Moreover, we assume that the states $|r_p\rangle$ are the complete set of
eigenstates with the eigenvalue $r_{\mathrm{max}}$. Our results obtained in
the previous sections reveal that this is true for a symmetric universal cloning
machine for any $d$ and for phase covariant cloning machine for $d=2,\ldots,7$.
Here we conjecture that this holds for phase covariant cloning machine and for
the Fourier cloner for arbitrary $d$.

If economical optimal cloning is possible, we must be able to construct
the pure states $|\Phi\rangle_{A,B,E}^l$ which appear in Eqs.
(\ref{lowcostoptimal}) and (\ref{lowcostoptimali})
as linear combinations of the states $|r_{p}\rangle$.
This means that there must exist
$d l_{\mathrm{max}}$ amplitudes $\alpha_{k}^l$
(with $\sum_{k=0}^{d-1} |\alpha_{k}^l|^2=1$) such that
\begin{equation}
\label{lowcostoptimalii}
U^l_{BE}\ket{k}_{B}\ket{\psi_{0}}_{E}=
\sum_{m,n,j=0}^{d-1} \alpha_{j+m}^l a_{m,n} \gamma^{n(k-j)}
|k+m\rangle_{B}|j \rangle_{E},
\end{equation}
for $l=1,\ldots, l_{\mathrm{max}}$.
The constraints (\ref{lowcostoptimalii}) are a necessary
condition for the existence of economical cloning, whenever the support
of the admissible CP maps $S^l$ associated to the economical cloning
transformations $U^l$ is spanned by the $d$ states
$|r_p\rangle$.

Let us assume that the optimal cloning state is given by Eq.~(\ref{ansatz})
with the amplitude matrix (\ref{Fouriercov'}), this includes
the whole class of symmetric and asymmetric universal and Fourier-covariant
cloning machines. If an economical realization of such cloners exists, then
there must exist $d$ amplitudes $\alpha_k$ and a unitary
transformation $U_{BE}$ which satisfy Eq. (\ref{lowcostoptimalii}).
On inserting the explicit formula (\ref{Fouriercov'}) for the
amplitude matrix $a_{m,n}$ into Eq. (\ref{lowcostoptimalii}) we obtain,
\begin{eqnarray}
U_{BE}\ket{k}_{B}\ket{\psi_{0}}_{E}=
\sum_{j,m=0}^{d-1} \alpha_{m+j}[x_{1}\delta_{m,0}+d x_{3}\delta_{j,k}
\nonumber\\
+x_{2}(d\delta_{m,0}\delta_{j,k}+1)]
\ket{m+k}_{B}\ket{j}_{E}.
\label{condi}
\end{eqnarray}
Unitarity (or equivalently trace preservation (\ref{tracepreservation}))
imposes the following condition:
\begin{equation}
\label{uni}
_{B}\bra{k^\prime} _{E}\bra{\psi_{0}}U_{BE}^+|
U_{BE}\ket{k}_{B}\ket{\psi_{0}}_{E}= \delta_{k,k'}
\end{equation}

Taking $k=k'$ in Eqs. (\ref{condi}) and (\ref{uni}) we get after some algebra
\begin{equation}
\sum_{j} |\alpha_j|^2 f_d(x_1,x_2,x_3)+ |\alpha_k|^2 g_d(x_1,x_2,x_3)=1,
\quad \forall k,
\label{diagonal}
\end{equation}
where $f_d$ and $g_d$ are second order polynomials in $x_j$,
\begin{eqnarray}
f_d&=&x_1^2+dx_2^2+d^2x_3^2+2x_1x_2+2dx_2x_3 \nonumber \\
g_d&=&(d^2+2d)x_2^2+2dx_1x_2+2dx_1x_3+2d^2x_2x_3. \nonumber \\
\end{eqnarray}
If we now consider the case $ k \neq k'$ in Eq. (\ref{uni}) we obtain
\begin{eqnarray}
&&(dx_2^2+2x_1x_2+2dx_2x_3) \sum_j \alpha_j \alpha_{j+k-k'}^\ast
\nonumber \\
&&+dx_2^2(\alpha_k\alpha_{2k-k'}^\ast
+\alpha_{2k'-k}\alpha_{k'}^\ast)+2dx_1x_3 \alpha_{k'}\alpha_k^\ast=0
\nonumber \\
\label{offdiagonal}
\end{eqnarray}
Normalization of the cloning state (\ref{ansatz}) imposes
that $f_d+g_d/d=1$. In virtue of Eq. (\ref{diagonal}),
either $|\alpha_{k}|^2=1/d$,
$\forall k$ or $g_d=0.$ The
latter constraint is neither satisfied by the universal nor by
the Fourier-covariant cloners so that in order
that such cloners admit an economical realization, $|\alpha_{k}|^2=1/d$
and the norms of all the $d$ a priori unknown parameters
$\alpha_{k}$ must be equal.
In order to ensure unitarity, it is still necessary to fulfill the condition
(\ref{offdiagonal}). It is worth noting that in (\ref{offdiagonal})
appear only products of $\alpha^*_{i}$ and $\alpha_{j}$ the indices
of which differ by the same quantity $i-j=k-k'$.
Hence, if we make the substitution $k'=k-m$ in Eq. (\ref{offdiagonal}) and then
sum over $m=0, \cdots, d-1$, we get the following constraint,
\[
g_d(x_1,x_2,x_3) \sum_{j}\alpha_j \alpha_{j+m}^\ast =0, \qquad m \neq 0.
\]
Since $g_d=0$ is never satisfied by the optimal universal and/or
Fourier-covariant cloners, we find that
$\sum_{j}\alpha_j \alpha_{j+m}^\ast =0$, $m \neq 0$.
As a consequence, the satisfaction of the condition (\ref{offdiagonal})
also implies
\begin{equation}
x_2^2(\alpha_k\alpha_{2k-k'}^\ast
+\alpha_{2k'-k}\alpha_{k'}^\ast)+2x_1x_3 \alpha_{k'}\alpha_k^\ast=0.
\end{equation}


In the case of the universal cloner, $x_2=0$ and $x_1x_3 \neq 0$
and it is clear that no solution exists for the system for any $d \geq 2$.

When the cloner is the optimal Fourier-covariant cloner,
it can be shown that the ancilla does not bring extra information about the
state under copy, which is expressed by the relation $x_{2}^2=x_{1}x_{3}$.
The amplitudes $\alpha_j$ must
then obey the relations
\begin{eqnarray}
\alpha_k\alpha_{2k-k'}^\ast
+\alpha_{2k'-k}\alpha_{k'}^\ast+2\alpha_{k'}\alpha_k^\ast&=&0,
\qquad \forall k\not= k'. \nonumber \\
\label{system}
\end{eqnarray}
We shall show that this system of equations admits a solution
only in dimension $d=2$. This solution corresponds to the (qubit) phase
covariant cloner and, in the symmetric case, to the symmetric Niu-Griffiths economical
realization already mentioned in a previous section
(see also Ref. \cite{lowcostcloning}). The asymmetric economical realization
was studied in detail in the reference \cite{lowcostcloning}.

Since all the amplitudes $\alpha_j$ have the same norm,
the triangular inequality together with Eq. (\ref{system}) implies that
\begin{equation}
\alpha_k
\alpha_{k+m}^\ast=\alpha_{k-2m}\alpha_{k-m}^{\ast}=-\alpha_{k-m}\alpha_{k}^\ast.
\label{fouriercondition}
\end{equation}
It is convenient to consider normalized amplitudes
$\tilde{\alpha}_j=\sqrt{d}\alpha_j$, $|\tilde{\alpha}_j|=1$.
Taking $m=1$ we obtain from Eq. (\ref{fouriercondition}) the
recurrence formula $\tilde{\alpha}_{k+1}=-\tilde{\alpha}_k^2
\tilde{\alpha}_{k-1}^\ast$. Without loss of generality, we can assume
$\tilde{\alpha}_0=1$ and express all $\tilde{\alpha}_j$ in terms of
$\tilde{\alpha}_1$ as follows: $\tilde{\alpha}_{2n}=(-1)^n \tilde{\alpha}_1^{2n}$
and $\tilde{\alpha}_{2n+1}=(-1)^n \tilde{\alpha}_1^{2n+1}$.
Substituting these expressions in the constraint
$\sum_{l=0}^{d-1}\alpha_{j}\alpha_{j+m}^*=0$ with $m=2$ leads to
$\tilde{\alpha}_1^2=0$, which contradicts the fact that
$|\tilde{\alpha}_1|=1$. It is only in dimension 2 that the contradiction can
be avoided because $m=2=0$ modulo $d$ in dimension 2.

The treatment of the phase-covariant cloner (symmetric and asymmetric as well)
presents many similarities with the treatment of the symmetric phase-covariant
cloner already discussed in the Section II, excepted that the parametrization is different:
when
the constraints \ref{phasecov} and \ref{lowcostoptimalii} are satisfied, it is
easy to derive the following system of equations:
\begin{eqnarray}
&&x_1^2+d^2x_3^2+|\alpha_{k}|^2 (g_d-2dx_2^2)=1, \qquad \forall k,\nonumber \\
&&2dx_1x_3\alpha^*_{-k'}\alpha_{k}=0; \qquad \forall k\not= k'
\label{system'}
\end{eqnarray}

The solution of the second constraint is $\alpha_{k}=\delta_{k,l}$. Inserting in the first constraint, we get the
equation $x_1^2+ d^2x_3^2=1$ which is fulfilled in dimension $d=2$ only,
in virtue of the identity $x_{3}^2=(x_{1}+x_{2}+x_{3})(x_{2}+x_{3})$, and of the
normalization of the cloning state $x_1^2+d^2x_3^2+dx_2^2+2x_1x_2+2x_1x_3 +2d.x_2x_3=1$,
in which case we recover the Fourier-covariant
cloner and its Niu-Griffiths economical realization already mentioned in the section 2. Note
that our proof constitutes a strong evidence of the impossibility of economical phase-covariant cloning in
any finite dimension different from 2, in agreement with the strict proof
of the section IIIB for dimensions 3 to 7. Note also that the 2 dimensional realization of the
phase-covariant cloner (\ref{phasecov}) differs, in our parametrization, from
the 2 dimensional Fourier covariant cloner (\ref{Fouriercov'}), but they can be shown
to be equivalent up to a change of basis and to a relabeling of the $x$ parameters.

\section{Conclusions}

In this paper, we have focused on
one-to-two cloning machines in arbitrary dimensions,
and have investigated the connections between the cloners with and without
ancillas. We have established a series of no-go theorems
for economical cloning, some of them being firm (universal cloner),
some others relying on an ansatz which was only tested numerically
(phase-covariant and Fourier-covariant cloners).
Note that, in our approach, the figure of merit
is the cloning fidelity, but it seems that the cloners
that optimize Eve's information also fulfill the ansatz (\ref{ansatz}).
In this case, the CP map approach is not very well adapted 
because of the non-linearity of the information measure. Nevertheless, 
we were able to establish the validity of the condition 
(\ref{lowcostoptimal}) in an independent manner, under the assumption of optimality of the ansatz only.

Our results strongly suggest that the Niu-Griffiths 
economical phase-covariant cloning machine for qubits is quite
unique among the optimal $1\rightarrow 2$ cloning machines. 
This conclusion is of importance
in connection with the security of quantum cryptographic protocols because
it shows that the realization of cloning attacks on quantum cryptographic
protocols that exploit higher-dimensional Hilbert spaces 
would require the mastering and control of three-qudit transformations,
which constitutes a serious technological challenge.
Another possibility would be of course to implement a sub-optimal 
economical phase-covariant cloner, as presented in Section IV, but
the resulting attack would be weaker.

To be complete, it is worth noting that in the limit of an infinite dimension,
the optimal phase-covariant, Fourier-covariant, and universal cloners
tend all to a fidelity of 1/2, 
for which an economical realization exists:
the original qudit is replaced by noise with probability 1/2 and directed
to Eve, or it is resent to Bob without disturbance while Eve gets noise.
In this rather trivial limit, economical cloning is always possible,
and extremely cheap!

In a future work, it would be interesting to study 
the possibility of economical one-to-N cloning, 
where it seems that the limitations are less drastic than in
the one-to-two case. For instance the one-to-three and one-to-N cloners
studied in \cite{123,12N,buscemi-economical}
also admit an economical realization.

\bigskip

\noindent {\it Note added:} A related work on economical quantum cloning
has been reported independently by the QUIT group in Pavia \cite{buscemi-economical}, following discussions we had during a visit
of the QUIT group in January 2004 which also led to the present paper.

\begin{acknowledgments}
Special thanks to Helle Bechmann-Pasquinucci (QUIT group, University of Pavia),
who is partially at the origin of this work, and to the other members of the
QUIT group for stimulating discussions.
TD is a Postdoctoral Fellow of the Fonds voor Wetenschappelijke
Onderzoek-Vlaanderen. This research was supported by the Belgian Office
for Scientific, Technical and Cultural Affairs in the framework of the
Inter-University Attraction Pole Program of the Belgian government
under grant V-18, the Fund for Scientific Research - Flanders
(FWO-V), the Concerted Research Action ``Photonics in Computing'',
the Solvay Institutes for Physics and Chemistry, and the research council (OZR) of the VUB. NJC and JF acknowledge financial support 
from the Communaut\'e Fran{\c c}aise de Belgique under grant ARC 00/05-251,
and from the EU under projects RESQ (IST-2001-37559) 
and CHIC (IST-2001-33578). JF also acknowledges support
from the grant LN00A015 of the Czech Ministry of Education.

\end{acknowledgments}

\end{document}